\documentclass[journal]{elsarticle}
\usepackage{graphicx}
\graphicspath{ {./plots/} }
\usepackage{amsmath}
\usepackage{color,soul}
\usepackage{subcaption}
\usepackage{caption}
\usepackage{multirow}
\usepackage{amsmath}
\usepackage{epstopdf}
\usepackage{xcolor}
\usepackage{booktabs}
\usepackage{lineno}
\usepackage{setspace}

\begin{document}

\begin{frontmatter}
\title{Neural network based terramechanics modeling and estimation for deformable terrains \tnoteref{funding}}

\author[1]{James Dallas}
\author[2]{Michael P. Cole}
\author[2]{Paramsothy Jayakumar}
\author[1]{Tulga Ersal \corref{cor}}

\cortext[cor]{Corresponding author: tersal@umich.edu}

 \maketitle

\address[1]{J. Dallas and T. Ersal are with the Department of Mechanical Engineering, University of Michigan, 1231 Beal Ave, Ann Arbor, MI 48109, USA.}
\address[2]{M.P. Cole and P. Jayakumar are with the U.S. Army Ground Vehicle Systems Center, 6501 E Eleven Mile Rd, Warren, MI 48397, USA.}

\tnotetext[funding]{This work was funded by U.S. Department of Defense under the prime contract number W56HZV-17-C-0005. DISTRIBUTION STATEMENT A.   Approved for public release; distribution unlimited. OPSEC \#3029}

\date{July 2019}

\begin{abstract}
	In this work, a neural network based terramechanics model and terrain estimator are presented with an outlook for optimal control applications such as model predictive control.  Recognizing the limitations of the state-of-the-art terramechanics models in terms of operating conditions, computational cost, and continuous differentiability for gradient-based optimization, an efficient and twice continuously differentiable terramechanics model is developed using neural networks for dynamic operations on deformable terrains.  It is demonstrated that the neural network terramechanics model is able to predict the lateral tire forces accurately and efficiently compared to the Soil Contact Model as a state-of-the-art model.  Furthermore, the neural network terramechanics model is implemented within a terrain estimator and it is shown that using this model the estimator converges within around 2\% of the true terrain parameter.  Finally, with model predictive control applications in mind, which typically rely on bicycle models for their predictions, it is demonstrated that utilizing the estimated terrain parameter can reduce prediction errors of a bicycle model by orders of magnitude.  The result is an efficient, dynamic, twice continuously differentiable terramechanics model and estimator that has inherent advantages for implementation in model predictive control as compared to previously established models.
\end{abstract}


\end{frontmatter}

\section{Introduction}
Autonomous ground vehicles (AGVs) have gained traction for military applications that could endanger human operators such as supply transport and reconnaissance \cite{Iagnemma2002}.  With regards to such applications, several considerations motivate this work.  First, vehicles are often required to operate on deformable terrains, where terrain properties are not explicitly known but greatly affect the vehicle's mobility \cite{Taheri2015, Dallas}.  Second, state-of-the-art autonomous navigation strategies often rely on model-dependent architectures \cite{liu2017,Liu2018}.  Third, efficient implementation of such navigation strategies require models to be twice continuously differentiable; however, state-of-the-art terramechanics models are often limited in terms of dynamic operation, computational complexity, or continuous differentiability \cite{liu2017, Taheri2015,Krenn2011,Ishigami2007,Smith2014,Guo2016, Dallas}. As such, in order to achieve safe and reliable operation of AGVs in off-road conditions, it is necessary to be able to learn a more accurate representation of the terrain online and capture this tire-terrain interaction through an efficient, dynamic, and twice continuously differentiable terramechanics model such that it can be implemented in planners to achieve more informed and reliable autonomous navigation.

Trajectory planning is a critical aspect in the development of autonomous vehicles.  It consists of sensing the environment a vehicle is operating in and determining control commands to safely navigate the vehicle in that environment \cite{liu2017}.  In this context, safety refers to not only defining a collision free path, but also avoiding dynamical safety hazards such as vehicle rollover.  Among the many methods available for safe navigation of autonomous vehicles, optimization based approaches are often preferred, as they allow for one to formally and explicitly implement safety constraints and vehicle dynamics while also remaining computationally efficient and ensuring optimality \cite{liu2017}.  However, inherent to the efficiency of such optimization based approaches, including Model Predictive Control (MPC), is the requirement that all functions in the optimal control problem be twice continuous differentiable \cite{liu2017}. For vehicles operating on deformable terrains, this requirement is restrictive due to the complexity of the interactions generated at the tire-terrain interface.  As such, it is important that the terramechanics model not only be of high fidelity under dynamic operation, such that the full operating range of the AGV can be realized, but also satisfy the constraint of twice continuous differentiability while remaining computationally efficient.

Terramechanics modeling can be divided into three general categories: (1) empirical models, which are the simplest, but face difficulties in application beyond the test conditions used in development; (2) physics based models, which have demonstrated the highest fidelity at the cost of high computational expense; and (3) semi-empirical models, which strike a balance between computational efficiency and fidelity, and hence are better suited for real-time estimation and control \cite{Taheri2015}. Of the semi-empirical methods, Bekker-based models have emerged as perhaps the most widely used \cite{Gallina2014,Ishigami2007,Smith2014,Guo2016}.  In these models the stresses are calculated over the contact patch between a rigid tire and the deformable terrain, and integrated to obtain the forces acting on the tire \cite{Smith2014}.  To accurately represent the complex stress distribution generated at the contact patch, Bekker-based models rely on numerous parameters that describe the terrain characteristics, such as cohesion and internal friction angle, to name a few.  However, knowledge of these parameters is limited in vehicle operation, where a vehicle may be operating on unknown terrains or terrains in which the properties vary.  Furthermore, classical Bekker-based models are often limited in application to steady-state operation \cite{Smith2014}.  

An extension of Bekker's method, known as the Soil Contact Model (SCM), essentially discretizes the tire-terrain interaction and allows for dynamic operation \cite{Taheri2015, Krenn2009}.  However, due to the discretization and integration of stress, SCM can potentially be too computationally expensive for real-time applications \cite{Dallas}.  In response to this limitation, a Bekker-based SCM surrogate model was developed to extend classical Bekker theory to account for some additional dynamic effects \cite{Dallas}.  While the surrogate model developed in \cite{Dallas} proved sufficient for estimation purposes, the lack of twice continuous differentiability  poses difficulties when utilized in model-dependent navigation algorithms, such as MPC \cite{liu2017}.  

As such, a computationally efficient, twice continuously differentiable dynamic terramechanics model for deformable terrains is still needed.  A potential candidate to address this need is neural networks, which have already demonstrated success in predicting tire forces for on-road applications \cite{Acosta2018,Kim1995,Matusko2008} and in slip detection on deformable terrains \cite{Gonzalez2018}; however, extending such approaches to lateral force prediction on deformable terrains is still an open research area.  The efficiency and continuous differentiability of neural networks makes them a suitable candidate for a terramechanics model in off-road model-dependent navigation architectures.  However, such surrogate models will still rely on numerous parameters that characterize the terrain properties that are likely to be unknown \textit{a priori} and hence need to be estimated online.

Researchers have already recognized the need for terrain estimation and several approaches can be found in the literature.  In \cite{Gallina2014,Gallina2016}, an in depth discussion of an offline Bayesian procedure for terrain parameter identification is presented.  
Others have proposed a linear least squares estimator for two terrain parameters: cohesion and internal friction angle \cite{Iagnemma2002,Iagnemma2004}.  However, the work relies on linearized terramechanics models, which can result in inaccurate force predictions \cite{ZhenzhongJia2011}, hence limiting its applicability in model-predictive architectures.  Finally, in \cite{Dallas} an accurate SCM surrogate model and an online terrain estimator has been proposed; however, the terramechanics model lacks the continuous differentiability required by many model-dependent navigation algorithms, such as MPC.  As such, an online terrain estimator that utilizes a twice 
continuously differentiable terramechanics model is needed.

Recognizing the needs identified above, this study presents a new approach for terramechanics modeling and its implementation within estimation for deformable terrains. First, due to the large computation time associated with integrating stresses in SCM and limitations of state-of-the-art terramechanics models, an efficient, dynamic, twice continuously differentiable terramechanics model is developed based on neural networks with sufficient agreement with SCM. Then, the neural network is implemented in the estimation architecture of \cite{Dallas} to identify the dominant terrain parameter, namely, the sinkage exponent.  The results are compared to that of \cite{Dallas}. The outcome is an efficient, dynamic, twice continuously differentiable terramechanics model and its implementation within an estimation architecture that can be used to better inform control and path-planning algorithms for AGVs.

The rest of the paper is organized as follows.  Sec. \ref{sec:TerramechanicsModels} briefly discusses several state-of-the-art terramechanics models for deformable terrains and presents the new neural-network based approach to terramechanics modeling, along with the terrain estimation architecture.  Sec. \ref{sec:VehicleModels} presents the vehicle models, i.e., the full-order model utilized in simulating the plant as well as the reduced-order model utilized in the estimation architecture.  Sec. \ref{sec:Results} demonstrates the ability of the proposed neural network architecture to estimate critical terrain parameters and improve model prediction capabilities as compared to that proposed in \cite{Dallas}.  Finally, Sec. \ref{sec:Conclusion} discusses the conclusions of this study.

\section{Terramechanics Modeling and Estimation} \label{sec:TerramechanicsModels}
\subsection{Terramechanics Models}\label{sec:terramech_description}

In this work, three terramechanics models are utilized for various purposes.  First, SCM is utilized as the ground truth and serves as the terramechanics model for the plant model described in Sec. \ref{sec:PlantModel}, as well as to generate training data for the neural network. However, SCM is rather complex and hence may not be suitable for online estimation \cite{Dallas}.  An overview of SCM can be found in \cite{Gallina2014,Krenn2011}.  Briefly, SCM relies on discretization of the terrain and based upon the deformations at each node, determines relevant stresses that are then integrated to obtain the tire contact force.  Second, a dynamic Bekker-based surrogate to SCM serves as a comparison between the neural network performance and state-of-the-art approaches.  More information on this model can be found in \cite{Dallas}.  Finally, due to the lack of twice continuous differentiability of this second model, a neural network is developed as the third terramechanics model and as one of the original contributions of the paper. This model is explained further next.

In this work, Latin hypercube sampling is utilized to generate a set of inputs for training a feedforward neural network.  Based upon the Bekker-based SCM surrogate described in \cite{Dallas}, the inputs considered for the neural network include the Bekker terrain properties, slip ratio, slip angle, tire velocity, load, and steering rate, because these variables have been demonstrated to impact the force generation at the tire terrain interface.  
\begin{table}
	\begin{center}
	\caption{Neural network input space.}
	\label{tab: 2}
	\begin{tabular}{cc}
		\toprule
		\textbf{Input}                  & \textbf{Range} \\
		Slip ratio            &  -1--1 (-)\\
		Slip angle          & -0.6--0.6 (rad)    \\   
        Longitudinal velocity          & 2--10 (m/s)    \\    
		Load          & 500--5500 (N)    \\    
		Steering rate          & -0.56--0.56 (rad/s)   \\  
		
		Soil deformation modulus          & 43000--2080000 ($\text{N}/\text{m}^{n+1}$)\\
			Sinkage exponent            &  0.3--1.3 (--)\\
			Shear deformation modulus&  0.01--0.024 (m)\\
			Cohesion            &  650--20700 (Pa)\\
			Internal friction angle          &  0.105--0.66 (rad)\\
		\bottomrule
	\end{tabular}
	\end{center}
\end{table}

A Latin hypercube sampling approach of the network inputs, with the ranges given in Table \ref{tab: 2}, is used in developing the data set for training the neural network.  The input ranges are determined from the vehicle states obtained from the simulations discussed in Sec. \ref{sec:Results}, while the terrain parameter space is compiled from the literature.  One difference in the training set is that an aggregate parameter is used for the soil deformation modulus, $k*$, as in \cite{Gallina2014}.  The network targets are generated by propagating the Latin hypercube samples through SCM.  Once the data is generated, the data set is split into 70\% training, 15\% validation,  and 15\% test sets.  Then, the MATLAB Deep Learning toolbox is used to train the network through Bayesian regularization backpropagation, a mean squared error performance function, and hyperbolic tangent sigmoid transfer functions that guarantee twice continuous differentiability. 50 neural networks are trained and the network with the best performance is selected as the surrogate terramechanics model.  Preliminary explorations of the network and training set size suggest that 3 layers of 35 neurons with approximately 10,000 Latin hypercube samples achieves sufficient performance for the purposes of terrain estimation.

\subsection{Terrain Estimation} \label{sec:TerrainEstimation}

In this work, the estimated terrain parameter is chosen to be the sinkage exponent, as Bekker-based terramechanics models have been shown to exhibit a higher sensitivity to it compared to other terrain parameters \cite{Gallina2014, Dallas}.  The remaining terrain parameters are set to nominal values based upon the specific terrain type, which can be obtained from terrain classification algorithms \cite{Howard2006,Weiss2008}.  To estimate the sinkage exponent, an unscented Kalman filter (UKF) is utilized as described in \cite{Dallas}, but with the difference that the neural network model described above is employed for estimation instead of the Bekker-based SCM surrogate of \cite{Dallas}.  Essentially, the UKF follows a predictor corrector scheme, where (1) predictions are performed  by a 3-DoF bicycle model appended with the sinkage exponent, and (2) correction is performed based upon measurements of the vehicle states. The UKF then utilizes the uncertainties associated with the 3-DoF bicycle model and measurements to determine the best estimate.  Further information on the general UKF description can be found in \cite{Wan,Kolas2009}.  While many other estimation techniques are available, the UKF is used in this work, because it was found to be a suitable balance between computational efficiency and accuracy \cite{Dallas}.


\section{Vehicle Models} \label{sec:VehicleModels}
In this work, two vehicle models are used: an 11-DoF model acting as the plant, and a simplified 3-DoF bicycle model that serves for vehicle predictions in the estimator. These models are summarized next.

\subsection{Plant Model} \label{sec:PlantModel} 
The physical vehicle in the simulation-based validation of the proposed surrogate model and terrain estimator is modeled through an 11-DoF notional military vehicle with SCM as the terramechanics model in Chrono software \cite{Chrono}.  The vehicle is composed of a double wishbone suspension, rack-pinion steering, 4-wheel drive, and a simple powertrain without a torque converter or transmission \cite{Dallas}.  Random additive Gaussian noise is then added to the states reported by the plant to simulate sensor noise and acts as the measurement in the UKF.  Table \ref{tab: 5} lists the standard deviations used in the noise model for each state.  This represents a worse-case scenario, as actual sensors typically offer lower noise levels \cite{Ryu2002}.

\begin{table}
	\begin{center}
		\caption{Measurement standard deviations used for sensor simulation.}
		\label{tab: 5}
		\begin{tabular}{cc}
			\toprule
			\bf State       & \bf Noise ($\sigma$)\\
			$x$           &1.2 (m)\\
			$y$     &1.2 (m) \\
			$\psi$          & 0.0175 (rad) \\
			$u$                 & 0.25 (m/s) \\
			$v$               &  0.25 (m/s)\\
			$\omega_z$                 &0.0175 (rad/s)  \\
			\bottomrule
		\end{tabular}
	\end{center}
\end{table}

\subsection{Bicycle Model}
A vehicle model is necessary for predicting the future vehicle states in the UKF of Sec. \ref{sec:TerrainEstimation}.  In this work, a 3-DoF bicycle model with forward Euler integration is utilized due to the model's ability to maintain a proper level of fidelity and efficiency for short-horizon predictions \cite{Liu2016, Dallas}.  The bicycle model is given as follows:
\begin{equation}
\dot{z_b} = \begin{bmatrix}
u\cos\psi-(v+L_\text{f}\omega_z)\sin\psi \\
u\sin\psi+(v+L_\text{f}\omega_z)\cos\psi \\
w_z \\
a_x \\
(F_\text{yf}+F_\text{yr})/M_t-u\omega_z \\
(F_\text{yf}L_\text{f}-F_\text{yr}L_\text{r})/I_{zz}
\end{bmatrix} \label{bicycle}
\end{equation}
where the state vector, $z_b$, is defined as
\begin{equation}
z_b :=
\begin{bmatrix}
x \\ y \\ \psi \\ u \\ v \\ \omega_z
\end{bmatrix}
=
\begin{bmatrix}
\textrm{global $x$  position  of  front  axle}\\
\textrm{global  $y$  position  of  front axle}\\
\textrm{yaw  angle}\\
\textrm{longitudinal  velocity} \\ 
\textrm{lateral  velocity }\\ 
\textrm{yaw  rate}
\end{bmatrix} \label{bicycle_states}
\end{equation}
and $M_t$ is the vehicle mass, $I_{zz}$ is the vehicle's yaw moment of inertia, and $L_\text{f}$ and $L_\text{r}$ are the distance from the vehicle's center of gravity to the front and rear axles, respectively.  Finally, $F_\text{yf}$ and $F_\text{yr}$ are the front and rear tire lateral forces acting on the vehicle, as reported from the neural network  terramechanics model.

\section{Results and Discussion} \label{sec:Results}

In this section, the performances of the neural network terramechanics model and the estimator are evaluated.  Evaluation is performed on two fronts: (1) the ability of the terramechanics model and estimator to accurately predict tire forces and the sinkage exponent, respectively, and (2) the impact these estimated parameters have on improving the prediction capability of the 3-DoF vehicle model.  The first evaluation provides insight into the performance of the neural network in predicting tire forces and estimating terrain parameters, while the second provides an assessment of the estimation algorithm's utility for future use in model predictive control schemes.  

\begin{table}
	\begin{center}
		\caption{Terrain parameters for the simulated terrain \cite{Smith2014}.}
		\label{tab:clay}
		\begin{tabular}{ccc}
			\toprule
			\bf Parameter            &\bf Symbol       &\bf Clay\\
	     	Cohesive modulus	       &$k_c$          & 13200 ($\text{N}/\text{m}^{n+1}$)\\
			Frictional modulus	       &$k_\phi$      &  692200 ($\text{N}/\text{m}^{n+2}$)\\
			Sinkage exponent	       &$n$            &  0.5 (--)\\
			Shear deformation modulus&$k$             &  0.01 (m)\\
			Cohesion                  &$c$             &  4140 (Pa)\\
			Internal friction angle  &$\phi$        &  0.2269 (rad)\\
			\bottomrule
		\end{tabular}
	\end{center}
\end{table}

To maintain consistency and a fair comparison with the terramechanics model proposed in \cite{Dallas}, the same clay Chrono simulation is used in the evaluation studies.  Briefly, the simulation is performed with the plant model of Sec. \ref{sec:PlantModel} operating on a clay-like SCM terrain.  The parameters used in representing the clay terrain are given in Table \ref{tab:clay}.  The vehicle is then subjected to sinusoidal steering commands, steering fully in both directions, and sinusoidal throttle commands such that non-constant speed is achieved.  Once the simulation completes, the data is corrupted with the noise described in Section \ref{sec:PlantModel} to simulate sensors.  More information on the simulation settings and the steering and velocity profiles can be found in \cite{Dallas}.

\begin{figure}
	\centering
	\includegraphics[width=3.5in]{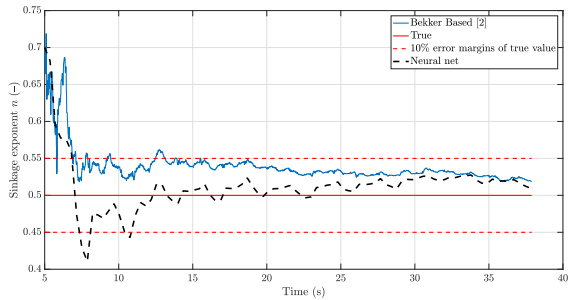}
	\caption{Simulated sinkage exponent estimation for neural network based terramechanics model (black dashed line) and Bekker-based model of \cite{Dallas} (blue solid line).}\label{fig:n_est}
\end{figure}

Table \ref{tab: est_results} summarizes the estimation results, including the initial guess, true sinkage exponent used in SCM, its estimated value and the percent error for the Bekker-based SCM surrogate of \cite{Dallas} and the neural network terramechanics model using the Chrono simulation on clay terrain.  Here, the estimated value is taken to be the converged value at the end of the simulation.  As can be seen, in both cases the estimator performs relatively well and estimates the parameter within 4\% of its true value, but the neural network reduces estimation error by nearly 50\% as compared to \cite{Dallas}.  It is also worth noting, as shown in Fig. \ref{fig:n_est}, that the neural network (black dashed line) appears to converge much faster to the true terrain parameter as compared to the model of \cite{Dallas} (blue solid line), which could prove beneficial in time critical applications, e.g. immediate obstacle avoidance. 
These findings suggest that the neural network is preferred for terrain estimation, since it can achieve a higher level of estimation accuracy as compared to the Bekker based model of \cite{Dallas} while also converging at a faster rate and having the beneficial property of twice continuous differentiability.

\begin{table}
	\begin{center}
		\caption{Performance comparison between neural network and Bekker-based SCM surrogate model in terms of estimated value of and estimation errors in the sinkage exponent $n$ on clay terrain. Bekker-based SCM surrogate results are from \cite{Dallas}.}
		\label{tab: est_results}
		\begin{tabular}{ccccc}
			\toprule
			\bf Model       & \bf Initial guess 			&\bf True val.	&\bf Est. val. &\bf \% error\\
			Bekker based \cite{Dallas}             &0.7 	&0.5 & 0.519  & 3.8\%\\
			Neural network             &0.7 	&0.5 & 0.5095  & 1.9\%\\
			\bottomrule
		\end{tabular}
	\end{center}
\end{table}

\begin{figure}
	\centering
		\includegraphics[width=3.3in]{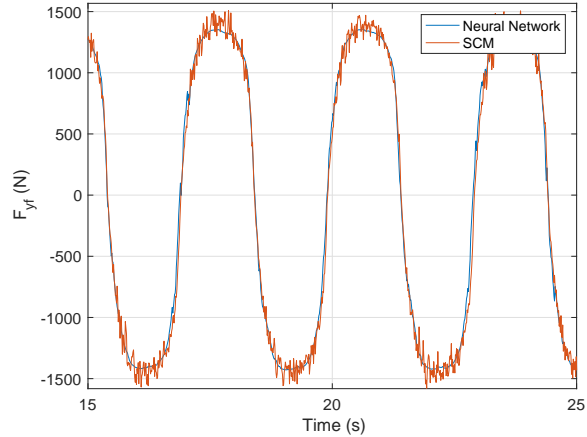}
	\caption{Simulated SCM and neural network lateral forces from estimation.}\label{fig: Fy}
\end{figure}

Fig. \ref{fig: Fy} shows the lateral forces from the front tire acting on the vehicle body given by SCM and the neural network running within the estimator.  As can be seen, the forces predicted by the neural network are reasonable as compared to SCM with a root-mean-squared-error of 91.76 N.  It should be noted that these forces are obtained from the sinusoidal vehicle simulation on clay, as discussed at the beginning of this section, and hence are completely different data than that generated by LHS in training the neural network.  As such, the good agreement observed in Fig. \ref{fig: Fy} suggests the network is able to generalize beyond its training and can potentially be applied to an MPC scheme.  

Furthermore, the peak computational time for a single UKF iteration of the neural network terramechanics model and estimator is 5.2 ms, whereas the peak computation time for the Bekker-based SCM surrogate is 10.5 ms on equivalent machines running MATLAB R2017a  \cite{Dallas}.  An optimized C++ version of the neural network and estimator has a peak computation time of 0.25 ms.  The estimator calls the bicycle model 17 times per UKF step, meaning the bicycle model, and neural network, can be evaluated efficiently and are conducive to real-time applications.  

As such, the results favor the accuracy, computational efficiency, and twice continuous differentiability of the neural network over the Bekker based model of \cite{Dallas}.

To assess the applicability of the proposed estimator and neural network for predictive applications, the bicycle model, with the neural network parameterized by the converged estimates, is used to predict the bicycle states approximately 2.5 s into the future.  This is chosen to mimic the procedure used by MPC and a full description is given in \cite{Dallas}.

\begin{table*}
	\begin{center}
		\caption{Mean squared error over entire simulation with 2.5 second prediction horizon for neural network and Bekker SCM surrogate.}
		\label{tab: pred_err}
		\begin{tabular}{ccccc}
			\toprule
			\bf Model &  \multicolumn{2}{c}{\bf Neural Network} & \multicolumn{2}{c}{\bf SCM Surrogate \cite{Dallas}} \\
			\midrule
			\bf State			&\bf n=0.5095  &\bf n=0.7&\bf n=0.519  &\bf n=0.7\\
			\midrule
			$x$           &0.037 (m)&0.031 (m)&0.037 (m)&0.01 (m)\\
			$y$     &0.071 (m) &0.334 (m)&0.022 (m)&0.15 (m) \\
			$\psi$          & 0.0018 (rad)& 0.023 (rad)&2.45e-04 (rad)& 0.0089 (rad) \\
			$u$                 & 1.33e-04(m/s) & 1.33e-04 (m/s)& 1.33e-04 (m/s)& 1.33e-04 (m/s)\\
			$v$               &  0.0065 (m/s)&0.285 (m/s)& 0.0047(m/s)& 0.15 (m/s)\\
			$\omega_z$                 &0.0015 (rad/s)&0.055 (rad/s)&9.01e-04 (rad/s)&0.023 (rad/s)\\ 
			\bottomrule
		\end{tabular}
	\end{center}
\end{table*}

Table \ref{tab: pred_err} gives the mean squared error (MSE) over the entire clay simulation with 2.5 second predictions for the bicycle model utilizing the neural network terramechanics model and the model reported in \cite{Dallas} for both the initial guess and converged value of the sinkage exponent. The baseline for this error calculation is the 11-DoF Chrono simulation.  As can be seen, utilizing the converged sinkage exponent for the neural network significantly reduces the MSE in the state prediction as compared to the initial guess, in some cases by an order of magnitude.  Comparing the state errors associated with the converged sinkage exponent for the neural network and the Bekker-based SCM surrogate, it can be seen that the errors of the neural network are slightly larger than that of the Bekker-based SCM surrogate; however, the errors for both models are quite close in general.  These results suggest that the neural network terramechanics model can be suitable for estimation and is better suited for application in control due to its comparable accuracy to the Bekker-based SCM surrogate and increased efficiency and twice continuous differentiability.  Furthermore, the reduction in prediction error achieved through the estimated parameters can potentially achieve better performing model predictive navigation and control.

\begin{figure}
	\centering
	\includegraphics[width=3.5in]{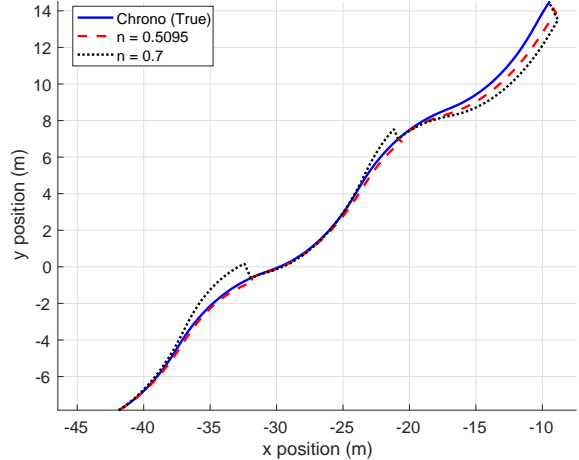}
	\caption{Chrono simulation for an AGV on SCM clay terrain with true vehicle positions from Chrono (blue solid line), bicycle model parameterized by $n = 0.5095$ (red dashed line), and bicycle model parameterized by initial terrain guess $n = 0.7$ (black dotted line) over an approximately 2.5 s prediction horizon. }\label{fig:traj}
\end{figure}

The increased performance of the bicycle model utilizing the estimated terrain parameter for the neural network is depicted in Fig. \ref{fig:traj}.  The position predictions of the bicycle model for the estimated terrain parameter $n = 0.5095$ (red dashed line) are much closer to the true Chrono simulation (blue solid line) as compared using the initial  guess $n = 0.7$ for the sinkage exponent (black dotted line).  It is anticipated that this improved position accuracy can be beneficial to autonomous obstacle avoidance and lane keeping tasks using MPC. Assessing these expected benefits systematically is subject to future work.

\section{Conclusion} \label{sec:Conclusion}
This paper considers AGVs operating on off-road deformable terrains and presents a novel neural network terramechanics model and its implementation within a terrain estimation algorithm.  The novelty is in the sense that the neural network is twice continuously differentiable, hence allowing for efficient implementation in MPC frameworks. Furthermore, the neural network achieves comparable accuracy to state-of-the-art dynamic terramechanics models while reducing the peak computation time.  The results suggest the neural network terramechanics model is able to predict tire lateral forces with sufficient accuracy for the problem of terrain estimation.  It is shown that the neural network is able to estimate the sinkage exponent with comparable accuracy to state-of-the-art terramechanics models while also satisfying the functional constraints of optimization based control (i.e., differentiability) and reduced computational time.  Finally, it is demonstrated that the estimated terrain parameters can significantly reduce the prediction errors of a 3-DoF bicycle model when the neural network terramechanics model is parameterized according to the estimated sinkage exponent.  Therefore, it is concluded that the neural network terramechanics model and its implementation within the estimator are an important advancement toward off-road AGVs.

Future work includes implementing the neural network model and estimation algorithm within MPC to assess the proposed architecture's utility.  It is also of interest to experimentally validate the neural-network-based estimation algorithm.

\bibliographystyle{elsarticle-harv}
\bibliography{terrain_est_publications,additional_refs,AMPC_bib,our_prior_work}

\end{document}